\documentclass[aps,prb,twocolumn]{revtex4}
\usepackage{amssymb}
\usepackage{graphicx}
\usepackage{hyperref}

\begin{document}
\title{Magnetic and thermal properties of the S~=~1/2 zig-zag spin-chain compound In$_2$VO$_5$}
\author{Yogesh Singh, R. W. McCallum, and D. C. Johnston}
\affiliation{Ames Laboratory and Department of Physics and Astronomy, Iowa State University, Ames, IA 50011}
\date{\today}

\begin{abstract}
Static magnetic susceptibility $\chi$, ac susceptibility $\chi_{\rm ac}$ and specific heat $C$ versus temperature $T$ measurements on polycrystalline samples of In$_2$VO$_5$ and $\chi$ and $C$ versus $T$ measurements on the isostructural, nonmagnetic compound In$_2$TiO$_5$ are reported.  A Curie-Wiess fit to the $\chi(T)$ data above 175~K for In$_2$VO$_5$ indicates ferromagnetic exchange between V$^{4+}$ (S~=~1/2) moments.  Below 150~K the $\chi(T)$ data deviate from the Curie-Weiss behavior but there is no signature of any long range magnetic order down to 1.8~K.  There is a cusp at 2.8~K in the zero field cooled (ZFC) $\chi(T)$ data measured in a magnetic field of 100~Oe and the ZFC and field cooled (FC) data show a bifurcation below this temperature.  The frequency dependence of the $\chi_{\rm ac}(T)$ data indicate that below 3~K the system is in a spin-glass state.  The difference $\Delta C$ between the heat capacity of In$_2$VO$_5$ and In$_2$TiO$_5$ shows a broad anomaly peaked at 130~K.  The entropy upto 300~K is more than what is expected for $S$~=~1/2 moments.  The anomaly in $\Delta C$ and the extra entropy suggests that there may be a structural change below 130~K in In$_2$VO$_5$.     

\end{abstract}

\maketitle

\section{Introduction}
\label{sec:INTRO}
Quasi-one dimensional spin-chain and spin-ladder materials have been studied partly in the hope of understanding the physics of the high $T_{\rm c}$ parent compounds\cite{bednorz1986} which can be viewed as an infinite array of spin-ladders in a plane, and also for the unusual quantum magnetism that these materials themselves exhibit.  For example, the spin-Peierls state in CuGeO$_3$ (S~=~1/2),\cite{hase1993} the spin gap in the two-leg ladder system SrCu$_2$O$_3$ (S~=~1/2),\cite{azmua1994} and the Haldane gap in Y$_2$BaNiO$_5$ (S~=~1) \cite{darriet1993} to name a few.  Holes doped into these quasi-two-dimensional ladders have also been predicted to pair and superconduct.\cite{dagotto1996}     
Frustration in addition to the low dimensionality leads to further exotic properties like the Wigner crystallization of magnons in the quasi-two-dimensional material SrCu$_2$(BO$_3$)$_2$,\cite{kodama2002} partial antiferromagnetic ordering in the spin chain materials Sr$_5$Rh$_4$O$_{12}$, Ca$_5$Ir$_3$O$_{12}$ and Ca$_4$IrO$_6$,\cite{cao2007} and order by disorder phenomena in the frustrated spin-chain compund Ca$_3$Co$_2$O$_6$.\cite{takeshita2006} 

\begin{figure}[t]
\includegraphics[width=3.5in]{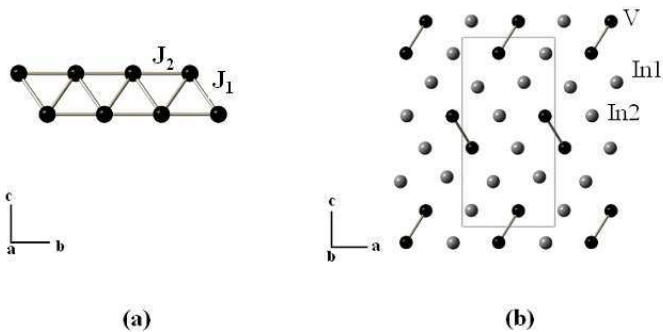}
\caption{(a) A segment of the crystal structure of In$_2$VO$_5$ along the $a$-axis showing the zig-zag chains of V$^{4+}$ moments along the $b$-axis with nearest neighbor ($J_1$) and next-nearest neighbor ($J_2$) exchange interactions.  (b) The crystal structure along the $b$-axis showing the arrangement of the chains in the $ac$-plane.  
\label{Figstructure}}
\end{figure}

Recently based on crystal chemical analysis, the orthorhombic ($Pnma$) compound In$_2$VO$_5$ has been proposed to be a $S$~=~$1/2$ zig-zag spin chain compound with competing nearest ($J_1$) and next nearest neighbor ($J_2$) antiferromagnetic exchange couplings between V$^{4+}$ moments.\cite{volkova2007}  The compound In$_2$VO$_5$ was first prepared both in polycrystalline and single crystalline form by Senegas and the single crystal structure was reported.\cite{senegas1975}  The crystal structure is shown in Fig.~\ref{Figstructure}.  Figure~\ref{Figstructure}(a) shows the zig-zag vanadium chains extending along the $b$-axis with nearest and next-nearest neighbor exchanges $J_1$ and $J_2$ respectively, and Fig.~\ref{Figstructure}(b) shows the structure along the $b$-axis showing the arrangement of the chains in the $ac$-plane.  The V atoms are crystallographically equivalent at room temperature.\cite{senegas1975}  A crystal chemical analysis estimated antiferromagnetic $J_1$ and $J_2$ with $J_2/J_1$~=~1.68\@.\cite{volkova2007}  Recent band structure calculations showed, however, that the room temperature crystal structure was incompatible with antiferromagnetic interactions and claimed that both $J_1$ and $J_2$ should be ferromagnetic instead.\cite{Schwingenschlogl2007}  Apart from the synthesis and crystal structure, an experimental study of the physical properties of this material have not been reported before.

Herein we report the magnetic and thermal properties of In$_2$VO$_5$ and the isostructural nonmagnetic compound In$_2$TiO$_5$.  The static magnetic susceptibility $\chi$, the ac susceptibility $\chi_{\rm ac}$ and the heat capacity $C$ versus temperature $T$ measurements on the compound In$_2$VO$_5$ and $\chi$ and $C$ versus $T$ measurements on the compound In$_2$TiO$_5$ are reported.  Our results indicate predominantly ferromagnetic exchange interactions between V$^{4+}$ moments above 150~K\@.  The product $\chi T$ versus $T$ shows a sudden reduction below about 120~K\@.  There is no signature of long-range magnetic ordering down to 1.8~K\@.  Surprisingly the $\chi(T)$ data, $\chi_{\rm ac}(T)$ at various frequencies and the $C(T)$ data suggest that stoichiometric In$_2$VO$_5$ undergoes a transition into a spin-glass state below about 3~K\@.  The difference $\Delta C$ between the heat capacities of In$_2$VO$_5$ and In$_2$TiO$_5$ shows a well-defined peak at 130~K and the entropy difference $\Delta S$ obtained by integrating the $\Delta C/T$ versus $T$ data up to 300~K is much larger than the value expected for $S$~=~1/2 moments.  The peak in $\Delta C$ and the extra entropy suggest a structural change in In$_2$VO$_5$ below $\sim$130~K\@.

\section{EXPERIMENTAL DETAILS}
\label{sec:EXPT}

Polycrystalline samples of In$_2$VO$_5$ and In$_2$TiO$_5$ were prepared by solid state synthesis.  The starting materials used were In$_2$O$_3$ (99.99\%, JMC), V$_2$O$_3$ (99.995\%, MV labs), V$_2$O$_5$ (99.995\%, MV Labs) and TiO$_2$ (99.999\%, MV labs).  Samples of In$_2$VO$_5$ were prepared by the reaction of In$_2$O$_3$ and VO$_2$ at 1100~$^\circ$C for 72~hrs in sealed quartz tubes with intermediate grindings after every 24~hrs.  The VO$_2$ was prepared by the reaction of equimolar quantities of the above V$_2$O$_3$ and V$_2$O$_5$ in a sealed quartz tube with an initial firing of 24~hrs at 900~$^\circ$C and a second firing of 24~hrs at 1000~$^\circ$C\@.  Samples of In$_2$TiO$_5$ were prepared by the reaction of stoichiometric quantities of In$_2$O$_3$ and TiO$_2$ at 1200~$^\circ$C  for 72~hrs in air with intermediate grindings after every 24~hrs.  
Hard well-sintered pellets were obtained.  Part of the pellet was crushed for powder x-ray diffraction (XRD).  The XRD patterns were obtained at room temperature using a Rigaku Geigerflex diffractometer with Cu K$\alpha$ radiation, in the 2$\theta$ range from 10 to 90$^\circ$ with a 0.02$^\circ$ step size. Intensity data were accumulated for 5~s per step.  The $\chi(T)$ and $\chi_{\rm ac}(T)$ were measured using a commercial Superconducting Quantum Interference Device (SQUID) magnetometer (MPMS5, Quantum Design) and the $C(T)$ was measured using a commercial Physical Property Measurement System (PPMS5, Quantum Design).  

\section{RESULTS}
\subsection{Structure of In$_2$VO$_5$ and In$_2$TiO$_5$}
\label{sec:RES-structure}

\begin{figure}[t]
\includegraphics[width=3.in]{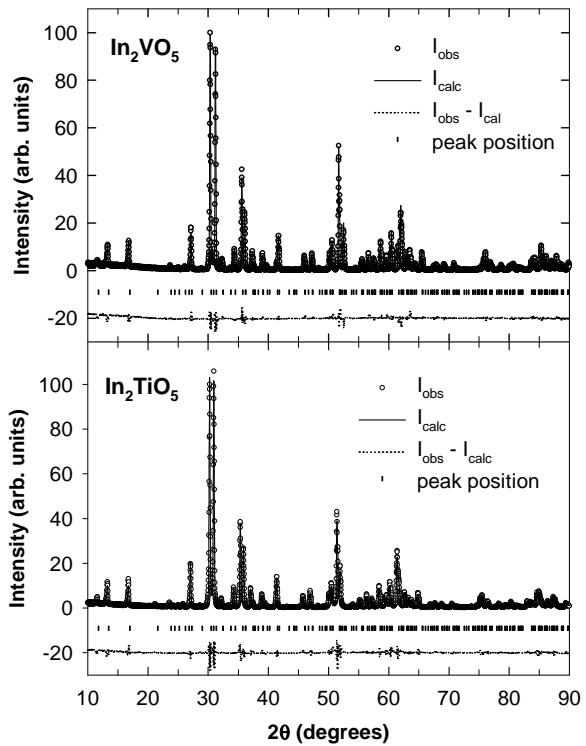}
\caption{Rietveld refinements of the In$_2$VO$_5$ and In$_2$TiO$_5$ X-ray diffraction data. The open symbols represent the observed data, the solid lines represent the fitted pattern, the dotted lines represent the difference between the observed and calculated intensities and the vertical bars represent the peak positions.  
\label{Figxrd}}
\end{figure}

\begin{table*}

\caption{\label{tabStruct}
Structure parameters for In$_2$VO$_5$ and In$_2$TiO$_5$ refined from powder XRD data.  The overall isotropic thermal parameter $B$ is defined within the temperature
factor of the intensity as $e^{-2B \sin^2 \theta/ \lambda^2}$.}
\begin{ruledtabular}
\begin{tabular}{l|c||cccccc}
Sample & atom & \emph{x} & \emph{y} & \emph{z} & $B$ &$R_{\rm wp}$&$R_{\rm p}$\\
  &  & & && (\AA$^2$) & \\\hline  
In$_2$VO$_5$ & In~~ & 0.0967(7) & 0.25 & 0.0851(2) & 0.003(1)& 0.224&0.154\\
& In~~ &0.3320(6) & 0.75 & 0.2374(2) & 0.006(1) & & \\
& V~~ &0.110(2) & 0.25 & 0.4194(7) & 0.013(3)& & \\
& O~~ &0.249(3) & 0.25 & 0.308(2) & 0.006(9)&  &\\  
& O~~ &0.339(3) & 0.25 & 0.502(2) & 0.009(9)& & \\
& O~~ &0.364(3) & 0.25 & 0.155(1) & 0.013(8)&  &\\ 
& O~~ &0.057(3) & 0.75 & 0.180(1) & 0.026(7)&  &\\ 
& O~~ &0.071(4) & 0.75 & 0.449(2) & 0.03(1)&  &\\ \hline  
In$_2$VO$_5$ & In~~ & 0.097(5) & 0.25 & 0.0852(2) & 0.008(1)& 0.195&0.139\\
& In~~ &0.3294(5) & 0.75 & 0.2389(2) & 0.006(1) & & \\
& Ti~~ &0.1098(12) & 0.25 & 0.4216(6) & 0.023(3)& & \\
& O~~ &0.245(3) & 0.25 & 0.315(1) & 0.002(6)&  &\\  
& O~~ &0.350(3) & 0.25 & 0.495(2) & 0.018(8)& & \\
& O~~ &0.362(2) & 0.25 & 0.156(2) & 0.03(1)&  &\\ 
& O~~ &0.061(3) & 0.75 & 0.180(1) & 0.016(9)&  &\\ 
& O~~ &0.066(3) & 0.75 & 0.447(2) & 0.035(9)&  &\\ 
\end{tabular}
\end{ruledtabular}
\end{table*}

All the lines in the X-ray patterns of In$_2$VO$_5$ and In$_2$TiO$_5$ could be indexed to the known\cite{senegas1975} orthorhombic \emph{Pnma} (No.~166) structure and Rietveld refinements,\cite{Rietveld} shown in Fig.~\ref{Figxrd}, of the X-ray patterns gave the lattice parameters $a$~=~~7.2282(5) \AA , $b$~=~3.4462(2) \AA\ and $c$~=~14.827(1) \AA\ for In$_2$VO$_5$, and $a$~=~~7.2419(9) \AA , $b$~=~3.5031(4) \AA\ and $c$~=~14.892(2) \AA\ for In$_2$TiO$_5$.  These values are in reasonable agreement with previously reported values for In$_2$VO$_5$ ($a$~=~~7.232 \AA , $b$~=~3.468 \AA\ and $c$~=~14.82 \AA ) and In$_2$TiO$_5$ ($a$~=~~7.237 \AA , $b$~=~3.429 \AA\ and $c$~=~14.86 \AA ).\cite{senegas1975}  Some parameters obtained from the Rietveld refinements of our samples are given in Table~\ref{tabStruct}.

\subsection{Magnetization Measurements}
\subsubsection{Isothermal Magnetization versus Magnetic Field}
\label{sec:MH}
\begin{figure}[t]
\includegraphics[width=3in]{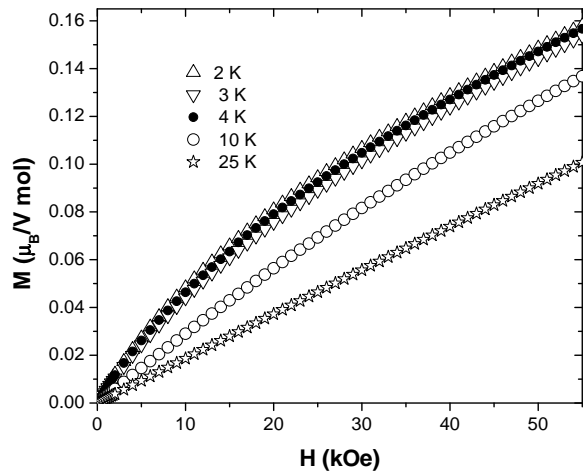}
\caption{ Isothermal magnetization $M$ versus magnetic field $H$ at various temperatures for In$_2$VO$_5$.
\label{FigMH}}
\end{figure}
\noindent
Figure~\ref{FigMH} shows the isothermal magnetization $M$ versus magnetic field $H$ measured at various temperatures $T$.  The $M(H)$ data at 25~K and at higher temperatures (not shown) are linear.  Below 10~K, however, the $M(H)$ data show a negative curvature at low fields with no sign of saturation even at the highest field $H$~=~55~kOe\@.  The $M(H)$ data could not be fitted by a Brillouin function assuming $S$~=~1/2 and $g$~=~2 or any other reasonable value of $S$ ($S$~=1 for example).  The reason can be seen from the data in Fig.~\ref{FigMH} for $T$~=~2, 3, and 4~K, which are nearly the same despite the factor of two range in temperature.  This indicates that the curvature in the $M(H)$ data is not due to the presence of nearly isolated paramagnetic impurities in the sample.  The negative curvature in Fig.~\ref{FigMH} may be associated with the spin-glass like state that we infer from our magnetic susceptibility and heat capacity data that we discuss below for In$_2$VO$_5$.

\subsubsection{Magnetic Susceptibility}
\label{sec:susceptibility}
\begin{figure}[t]
\includegraphics[width=3in]{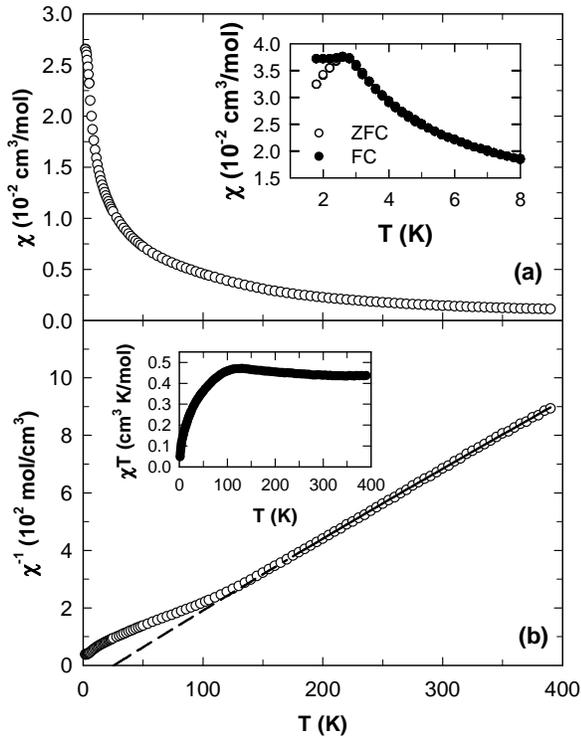}
\caption{(a) The magnetic susceptibility $\chi$ versus temperature $T$ for In$_2$VO$_5$ measured with an applied magnetic field of 1~T\@.  The inset in (a) shows the zero field cooled (ZFC) and field cooled (FC) $\chi (T)$ data measured in an applied magnetic field of 100~Oe\@.  (b) The inverse susceptibility $\chi^{-1}$ versus temperature $T$ data (symbols) and a fit by a Curie-Weiss model (solid line) extrapolated to lower temperatures (dashed line).  The inset in (b) shows the $\chi T$ versus $T$ data revealing a sudden reduction below about 120~K\@. 
\label{Figsus}}
\end{figure}
\noindent
\begin{figure}[t]
\includegraphics[width=3in]{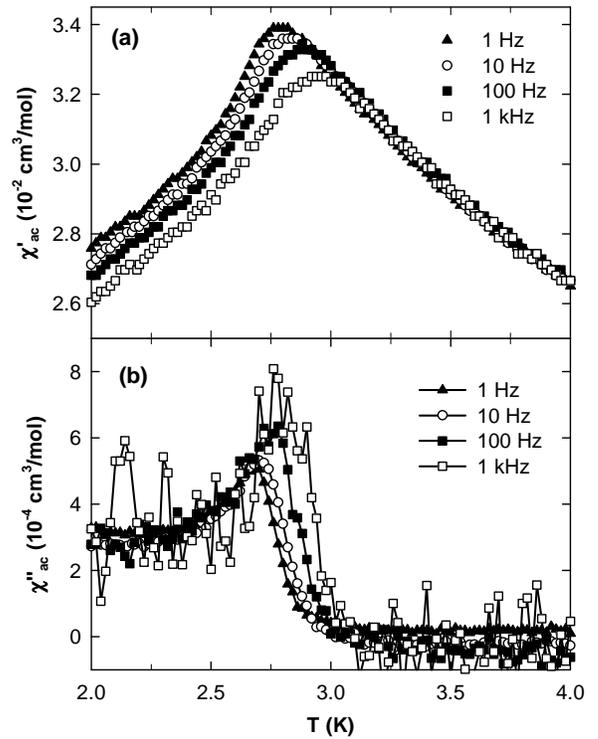}
\caption{(a) The real part of the ac magnetic susceptibility $\chi'_{\rm ac}$ versus temperature $T$ data between 2~K and 4~K for In$_2$VO$_5$ measured with various drive frequencies.  (b) The imaginary part of the ac magnetic susceptibility $\chi''_{\rm ac}$ data shown in (a) versus temperature $T$.   
\label{Fig-acsus}}
\end{figure}
The temperature dependence of the magnetic susceptibility $\chi \equiv M/H$ between 1.8~K and 400~K for In$_2$VO$_5$ measured in an applied magnetic field $H$~=~1~T is shown in the main panel of Fig.~\ref{Figsus}(a).  The inset in Fig.~\ref{Figsus}(a) shows the zero-field-cooled (ZFC) and field-cooled (FC) $\chi$ data between 1.8~K and 8~K measured in an applied magnetic field $H$~=~100~Oe\@.  The ZFC data show a cusp at 2.8~K and the ZFC data and FC data bifurcate below this temperature.  This feature is suggestive of a spin-glass state and we will return to these data when we discuss the ac susceptibility data below.  Figure~\ref{Figsus}(b) shows the $\chi^{-1}(T)$ data between 1.8~K and 400~K.  The $\chi(T)$ data between 175~K and 400~K were fitted by the Curie-Weiss expression $\chi = \chi_0+C/(T-\theta$) and the fit, shown as the solid curve through the $\chi^{-1}(T)$ data in Fig.~\ref{Figsus}(b) and extrapolated to lower temperatures (dashed curve), gave $\chi_0$~=~$-5.5(8)$$\times 10^{-5}$~cm$^3$/mol, $C$~=~0.386(7)~cm$^3$~K/mol and $\theta$~=~$26(2)$~K\@.  The value of $C$ corresponds to an effective moment of 1.75(2)~$\mu_{\rm B}$/V which is close to the value (1.73~$\mu_{\rm B}$/V) expected for $S$~=~1/2 moments with a $g$-factor of 2\@.  The positive value for $\theta$ indicates that the dominant interactions between the V$^{4+}$ moments are ferromagnetic.  This supports the results of the band structure calculations.\cite{Schwingenschlogl2007}  The $\chi^{-1}(T)$ data deviate from the Curie-Weiss behavior below 150~K, showing an upward curvature.  The $\chi^{-1}(T)$ decreases again below about 5~K which corresponds to the temperature of the cusp in the $\chi$ data in the inset of Fig.~\ref{Figsus}(a).  The inset in Fig.~\ref{Figsus}(b) shows the $\chi T$ versus $T$ data showing a pronounced reduction below about 120~K where the $\chi(T)$ data deviate from the Curie-Weiss law.  

Let us now return to the low field $\chi(T)$ data in the inset of Fig.~\ref{Figsus}(a).  The cusp in the ZFC $\chi$ data and the bifurcation between the ZFC and FC susceptibility suggests spin-glass freezing with a freezing temperature $T_{\rm f}$~=~2.8~K\@.  To check this possibility we have measured the ac susceptibility $\chi_{\rm ac}$ at various frequencies around the temperature of the cusp.  The real $\chi'_{\rm ac}$ and imaginary $\chi''_{\rm ac}$ parts of the $\chi_{\rm ac}$ data between 2~K and 4~K, measured in an ac field of 1~Oe and with various drive frequencies, are shown in Figs.~\ref{Fig-acsus}(a) and (b).  The maximum in the $\chi'_{\rm ac}$ data at 1~Hz occurs at 2.76~K and monotonically moves up in temperature with increasing frequency.  This trend is also seen in the $\chi''_{\rm ac}$ data.  The low field $\chi$ data in the inset of Fig.~\ref{Figsus}(a) and the $\chi_{\rm ac}$ data in Figs.~\ref{Fig-acsus}(a) and (b), together with the absence of a noticeable anomaly at $T_{\rm f}$ in the heat capacity data discussed below, are features usually observed at a spin-glass transition\cite{fisher} and strongly point to the existence of a spin-glass state below $T_{\rm f}$~=~2.8~K in In$_2$VO$_5$.

\begin{figure}[t]
\includegraphics[width=3in]{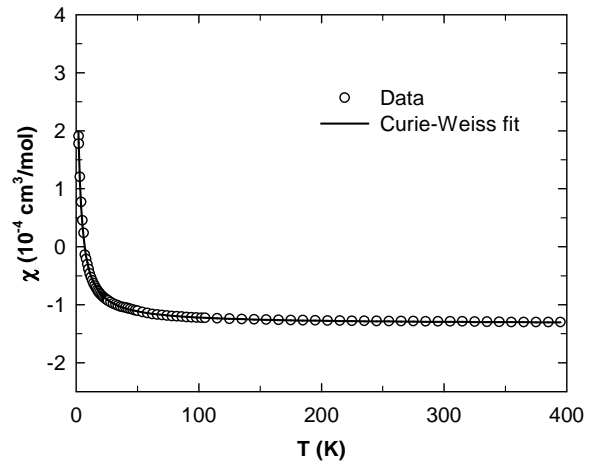}
\caption{ The magnetic susceptibility $\chi$ versus temperature $T$ for In$_2$TiO$_5$ measured with an applied magnetic field of 2~T\@.  the solid curve through the data is a fit by the Curie-weiss law.  
\label{Figsus2}}
\end{figure}

The magnetic susceptibility $\chi$ versus temperature $T$ for the isostructural and nominally nonmagnetic compound In$_2$TiO$_5$ is shown as open symbols in Fig.~\ref{Figsus2}.  The $\chi(T)$ is negative and almost temperature independent in the whole temperature range except for a small upturn at low temperatures.  The $\chi(T)$ data in the whole temperature range was fitted by a Curie-Wiess expression $\chi = \chi_0 + C/(T-\theta)$.  The fit, shown in Fig.~\ref{Figsus2} as the solid curve through the data, gave the values $\chi_0$~=~$-1.33(1)\times10^{-4}$~cm$^3$/mol, $C$~=~8.13(2)$\times$10$^{-4}$~cm$^3$/mol (corresponding to about 4~mol\% spin-1/2 impurities) and $\theta$~=~$-$1.60(5)~K\@.  The value of $\chi_0$ is of the same order as the core susceptibility $\chi_{\rm core} =-1.0 \times 10^{-4}$~cm$^3$/mol obtained from the sum of the atomic diamagnetic susceptibilities\cite{core} of the atoms in the material.   

\noindent

\subsection{Heat Capacity}
\label{sec:RES-heatcapacity}
\begin{figure}[t]
\includegraphics[width=3in]{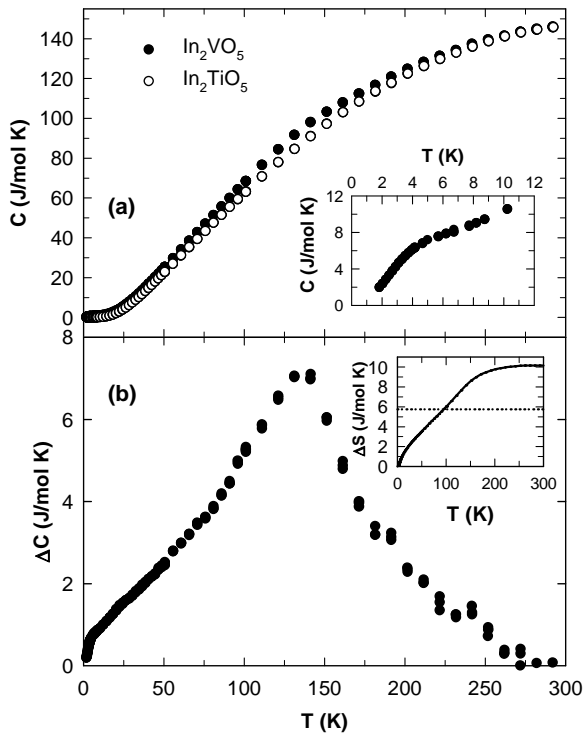}
\caption{(a) The heat capacity $C$ versus temperature $T$ for In$_2$VO$_5$ and In$_2$TiO$_5$\@.  The inset in (a) shows the $C$ data between 1.8~K and 10~K to highlight the anomaly at 2.8~K.  (b) The difference $\Delta C$ between the heat capacities of In$_2$VO$_5$ and In$_2$TiO$_5$\@.  The inset in (b) shows the difference entropy $\Delta S$ (solid curve) versus $T$ obtained by integrating $\Delta C/T$ versus $T$.  The dotted line corressponds to the value $\Delta S$~=~Rln2~=~5.67~J/mol~K expected for $S$~=~1/2 moments.    
\label{FigHC}}
\end{figure}

The heat capacity $C$ versus temperature $T$ of In$_2$VO$_5$ and In$_2$TiO$_5$ between 1.8~K and 300~K is shown in Fig.~\ref{FigHC}(a).  There is no obvious anomaly in the $C(T)$ data for In$_2$VO$_5$ either at 150~K where the deviation in $\chi(T)$ from the Curie-Weiss behavior was observed, or at 2.8~K where the cusp in the low field $\chi$ data was observed.  The inset in Fig.~\ref{FigHC}(a) shows the $C$ data for In$_2$VO$_5$ between 1.8~K and 10~K showing the weak shoulder with a maximum at about 5~K\@.  A broad anomaly with a maximum at a slightly higher temperature than the temperature $T_{\rm f}$ of the cusp observed in the $\chi$ data is also a feature observed for a spin-glass transition.\cite{fisher}  Figure~\ref{FigHC}(b) shows $\Delta C$ versus $T$, obtained by subtracting the heat capacity of In$_2$TiO$_5$ from that of In$_2$VO$_5$.  The $\Delta C$ shows a broad anomaly with a well defined maximum at about 130~K\@.  A significant $\Delta C$ spread over the whole temperature range suggests the presence of short range magnetic interactions.  However, the relatively sharp feature in $\Delta C$ at $\approx$~130~K is in contrast to the gradual change in the $\chi(T)$ data below this temperature.  This indicates that the anomaly in $\Delta C$ is at least not entirely of magnetic origin.  The inset in Fig.~\ref{FigHC}(b) shows the difference entropy $\Delta S$ versus $T$ (solid curve) obtained by integrating the $\Delta C/T$ versus $T$ data.  The horizontal dotted line corresponds to the value Rln2~=~5.67~J/mol~K expected for disordered $S$~=~1/2 moments.  The $\Delta S$ at 300~K exceeds Rln2 by 75\% which indicates that the whole $\Delta C(T)$ cannot be accounted for just by magnetism, and that the anomaly at 130~K in $\Delta C$ arises from a lattice change of some kind.  
  
The $C(T)$ data of the nonmagnetic compound In$_2$TiO$_5$ between 1.8~K and 10~K could be fitted by the expression $C$~=~$\beta T^3$ for the lattice heat capacity and gave the value $\beta$~=~0.100(1)~mJ/mol K$^4$.  From the value of $\beta$ one can obtain the Debye temperature $\theta_{\rm D}$ using the expression \cite{Kittel}
\begin{equation}
\Theta_{\rm D}~=~\bigg({12\pi^4{\rm R} n \over 5\beta}\bigg)^{1/3}~, 
\label{EqDebyetemp}
\end{equation}
\noindent
where R is the molar gas constant and $n$ is the number of atoms per formula unit (\emph{n}~=~8 for In$_2$TiO$_5$).  We obtain $\Theta_{\rm D}$~=~539(2)~K for In$_2$TiO$_5$.  The low temperature $C(T)$ of In$_2$VO$_5$ could not be used to obtain $\theta_{\rm D}$ because of the presence of the shoulder at 5~K\@.  

\noindent

\section{Discussion}
Our data have revealed unusual magnetic and thermal properties of the $S$~=~1/2 zig-zag spin-chain compound In$_2$VO$_5$.  High temperature $\chi(T)$ data indicate that there are predominantly ferromagnetic interactions between the V$^{4+}$ moments.  However, there is no signature of long-range magnetic ordering for the compound down to 1.8~K\@.  Instead a spin-glass state is indicated below 3~K which in turn suggests the presence of atomic or magnetic disorder and frustrating interactions in the system at this temperature.  The $S$~=1/2 zig-zag spin-chain model with ferromagnetic and antiferromagnetic exchange interactions $J_1$ and $J_2$ has recently been investigated and the susceptibility versus temperature, magnetization versus magnetic field and heat capacity versus temperature are reported for various values of $J_1$ and $J_2$.\cite{meisner2006, lu2006}  Our $\chi(T)$, $M(H)$ and $C(T)$ data cannot be fitted consistently by these models for any value of the parameters $J_1$ and $J_2$.  For example, from Fig.~3 in Ref.~\onlinecite{lu2006}, for the case of ferromagnetic nearest-neighbor interaction $J_1$ and antiferromagnetic next-nearest-neighbor interaction $J_2$, a maximum in $\chi$ is predicted at $T/J_1$~=~0.02\@ for $J_2$~=~$-0.28J_1$ to $-0.4J_1$.  The $\chi$ data for In$_2$VO$_5$ in Fig.~\ref{Figsus} can be qualitatively explained by the prediction with $J_1$~=100--150~K\@.  However, the $C(T)$ and $C(T)/T$ calculated in Ref.~\onlinecite{lu2006} for these values of $J_1$ and $J_2$ show a behavior at low temperatures which is not observed for In$_2$VO$_5$ in Fig.~\ref{FigHC}.  The $M(H)$ calculated in Ref.~\onlinecite{lu2006} shows a positive curvature at low $H$ before saturating at high $H$ which is also qualitatively different from the $M(H)$ that we observe in Fig.~\ref{FigMH}.  

The well defined anomaly in $\Delta C(T)$ at 130~K in Fig.~\ref{FigHC} and the large entropy difference $\Delta S$ above expectation for $S$~=~1/2 moments together suggest a structural change, a strong magnetoelastic coupling and/or different lattice dynamics of the two materials In$_2$VO$_5$ and In$_2$TiO$_5$ near this temperature.  

\section{CONCLUSION}
\label{sec:CON}
We have synthesized polycrystalline samples of the compounds In$_2$VO$_5$ and In$_2$TiO$_5$ and studied their structural, magnetic and thermal properties.  Our results for In$_2$VO$_5$ indicate predominantly ferromagnetic exchange interactions between V$^{4+}$ moments above 150~K\@.  There is no signature of long-range magnetic ordering down to 1.8~K\@.  The $\chi(T)$ data, $\chi_{\rm ac}(T)$ at various frequencies and the $C(T)$ data all suggest that In$_2$VO$_5$ undergoes a transition into a spin-glass state below about 3~K\@.  This in turn suggests the presence of lattice and magnetic disorder and frustration in the material.  The difference $\Delta C$ between the heat capacities of In$_2$VO$_5$ and In$_2$TiO$_5$ shows a well defined peak at 130~K and the entropy difference $\Delta S$ obtained by integrating the $\Delta C/T$ versus $T$ data up to 300~K is much larger than the value expected for $S$~=~1/2 moments.  The peak in $\Delta C$ and the extra entropy suggest that a structural change may occur in In$_2$VO$_5$ below 130~K, which in turn could change the V-V interactions. \\\\

\noindent
{\bf Note added:} After completion of this work, two preprints appeared \cite{Muller2007, simon2007} on the experimental study of In$_2$VO$_5$ and both In$_2$VO$_5$ and In$_2$TiO$_5$ respectively.  Reference \onlinecite{Muller2007} reports on the structural, magnetization, electrical resistivity and nuclear- and electron spin resonance measurements on In$_2$VO$_5$.  The value of the magnetic susceptibility $\chi$ at 1.8~K for In$_2$VO$_5$ reported in this preprint is about half of what we observe for our sample.  Low temperature X-ray data show an anomalous expansion in the $a$ and $b$ lattice parameters below 120~K\@.  Their $^{51}$V NMR measurements show a peak at 20~K in the nuclear relaxation rate $1/T_1$ versus $T$ data which they ascribe to a slowing down of spin fluctuations below this temperature.  They do not observe any signature in their measurments of the spin-glass state around 3~K as we infer from our data.  Reference \onlinecite{simon2007} reports on the magnetic susceptibility, heat
capacity and synchrotron powder x-ray diffraction measurements on In$_2$VO$_5$ and heat capacity measurements on In$_2$TiO$_5$.  The reported value of $\chi$ for In$_2$VO$_5$ at 1.8~K is consistent with our results.  They observe a peak in their $\chi$ data below 3~K and ascribe it to freezing of spin dimers into a singlet state below the temperture of the maximum in $\chi$.  Their low temperature synchrotron X-ray data show an anomalous expansion in the $b$ lattice parameters below 125~K\@.

The qualitative behavior of the $\chi$ reported in Refs.~\onlinecite{Muller2007,simon2007} is consistent with our observations.  The well-defined anomaly in the $\Delta C(T)$ that we observe at 130~K is in contrast with the subtle (less than 0.5\%) variation of the lattice parameters below this temperature reported in these two preprints.  Evidence of the spin-glass state below 3~K in our data is not reported in these preprints.  We do not think that the ground state of In$_2$VO$_5$ is a spin-singlet as suggested in Ref.~\onlinecite{simon2007}, because our data indicate a spin-glass ground state.
 
A consistent understanding of the physical properties of In$_2$VO$_5$ has not yet been reached and further experimental study will be required.

\begin{acknowledgments}
Work at the Ames Laboratory was supported by the Department of Energy-Basic Energy Sciences under Contract No.\ DE-AC02-07CH11358.  
\end{acknowledgments}

\end{document}